\documentstyle[editedvolume]{crckapb}  

\begin{opening}
\title{Radio Pulsars --- An Observer's Perspective}
\author{D.R. Lorimer}
\institute{National Astronomy and Ionospheric Center\\
           Arecibo Observatory\\
	   HC3 Box 53995, 
           Arecibo PR 00612, USA}
\end{opening}
\runningtitle{Radio Pulsars}

\begin{document}
\begin{figure}[hbt]
\setlength{\unitlength}{1in}
\begin{picture}(0,0)
\put(0,2.6){\includegraphics{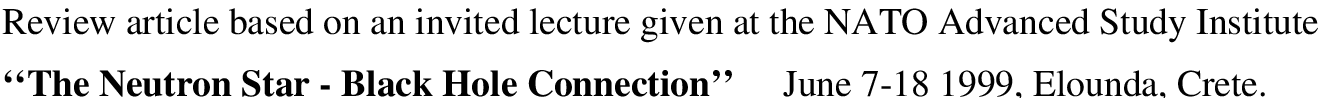}}
\end{picture}
\end{figure}
\begin{abstract}
Pulsar astronomy is currently enjoying one of the most productive phases 
in its history. In this review, I outline some of the basic observational 
aspects and summarise some of the latest results of searches for pulsars 
in the disk of our Galaxy and its globular cluster system.
\end{abstract}

\section{Preamble}

Pulsar astronomy began serendipitously in 1967 when Jocelyn Bell and
Antony Hewish discovered periodic signals originating from distinct
parts of the sky via pen-chart recordings taken during an
interplanetary scintillation survey of the radio sky at 81.5 MHz
(Hewish et al.~1968). This remarkable phenomenon has since been
unequivocally linked with the radiation produced by a rotating neutron
star (Gold 1968, Pacini 1968). Baade \& Zwicky (1934) were the first
to hypothesise the existence of neutron stars as a stable
configuration of degenerate neutrons formed from the collapsed remains
of a massive star after it has exploded as a supernova.

Although the theory of pulsar emission is complex, the basic idea can
be simply stated as follows: as a neutron star rotates, charged
particles are accelerated out along its magnetic poles and emit
electromagnetic radiation. The combination of rotation and the beaming
of particles along the magnetic field lines means that a distant
observer records a pulse of emission each time the magnetic axis
crosses his/her line of sight, i.e.~one pulse per stellar
rotation. Like many things in life, the emission does not come for
free. It takes place at the expense of the neutron star's rotational
kinetic energy --- one of the key predictions of the Gold/Pacini
theory. Measurements of the secular increase in pulse period through
pulsar timing techniques (\S \ref{sec:timing}) are in excellent
agreement with this idea.

\begin{figure}[hbt]
\setlength{\unitlength}{1in}
\begin{picture}(0,3)
\put(0.75,-0.05){\includegraphics{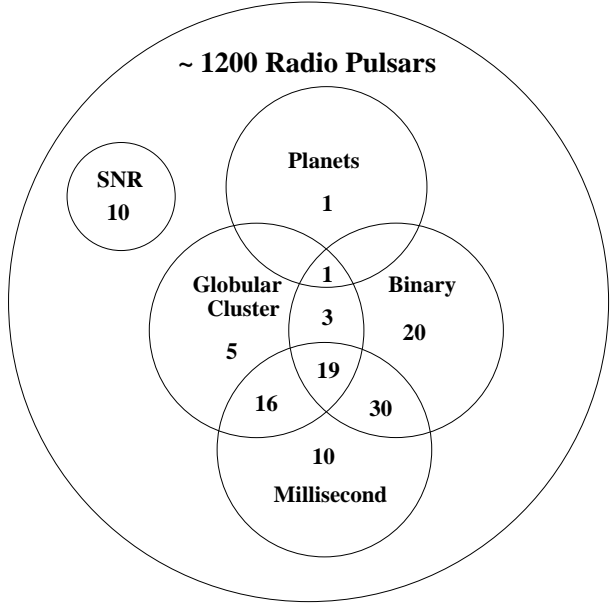}}
\end{picture}
\caption[]
{
An adaptation of Dick Manchester's Venn diagram showing the various types
of radio pulsars.  SNR denotes pulsars likely to be associated with
supernova remnants.
}
\label{fig:venn}
\end{figure}

Pulsar astronomy has come a long way in a remarkably short space of
time. Systematic surveys with the world's largest radio telescopes over
the last 30 years have revealed more than 1200 pulsars in a rich variety of
astrophysical settings. The present ``zoo'' of objects is summarised
in Fig.~\ref{fig:venn}.
Some of the highlights so far include: (1) The original {\sl binary
pulsar} B1913+16 (Hulse \& Taylor 1975) --- a pair of neutron stars in
a 7.75-hr eccentric orbit.  The measurement of the orbital decay due
to gravitational radiation from the binary system resulted in the 1993
Physics Nobel Prize.  (2) The original {\sl millisecond pulsar},
B1937+21, discovered by Backer et al.~(1982) has a period of only
1.5578 ms. The implied spin frequency of 642 Hz means that this
neutron star is close to being torn apart by centrifugal forces. (3)
Pulsars with planetary companions. Well before the discoveries of
Jupiter-mass planets by optical astronomers, Wolszczan \& Frail (1992)
discovered the millisecond pulsar B1257+12 accompanied by three
Earth-mass planets --- the first planets discovered outside our Solar
system. (4) Pulsars in globular clusters.  Dense globular clusters are
unique breeding grounds for exotic binary systems and millisecond
pulsars.  Discoveries of 40 or so ``cluster pulsars'' have permitted
detailed studies of pulsar dynamics in the cluster potential, and of
the cluster mass distribution.

The plan for the rest of this review is as follows: \S \ref{sec:dm}
covers pulse dispersion and what it tells us about pulsars and the
interstellar medium; \S \ref{sec:profs} discusses pulse profiles and
their implications. In \S \ref{sec:timing} the essential aspects of
pulsar timing observations are reviewed. \S \ref{sec:searching}
reviews the techniques employed by pulsar searchers. Finally, in \S
\ref{sec:recent}, we summarise some of the recent results from
searches at Parkes. More complete discussions on the observational
aspects covered here can be found in Lyne \& Smith (1998).

\section{Pulse Dispersion and the Interstellar Medium}
\label{sec:dm}

Newcomers to pulsar astronomy would do well to begin their studies
by reading the discovery paper (Hewish et al.~1968), a classic article
packed with observational facts and their implications.

\begin{figure}[hbt]
\setlength{\unitlength}{1in}
\begin{picture}(0,3.2)
\put(0.5,0){\includegraphics{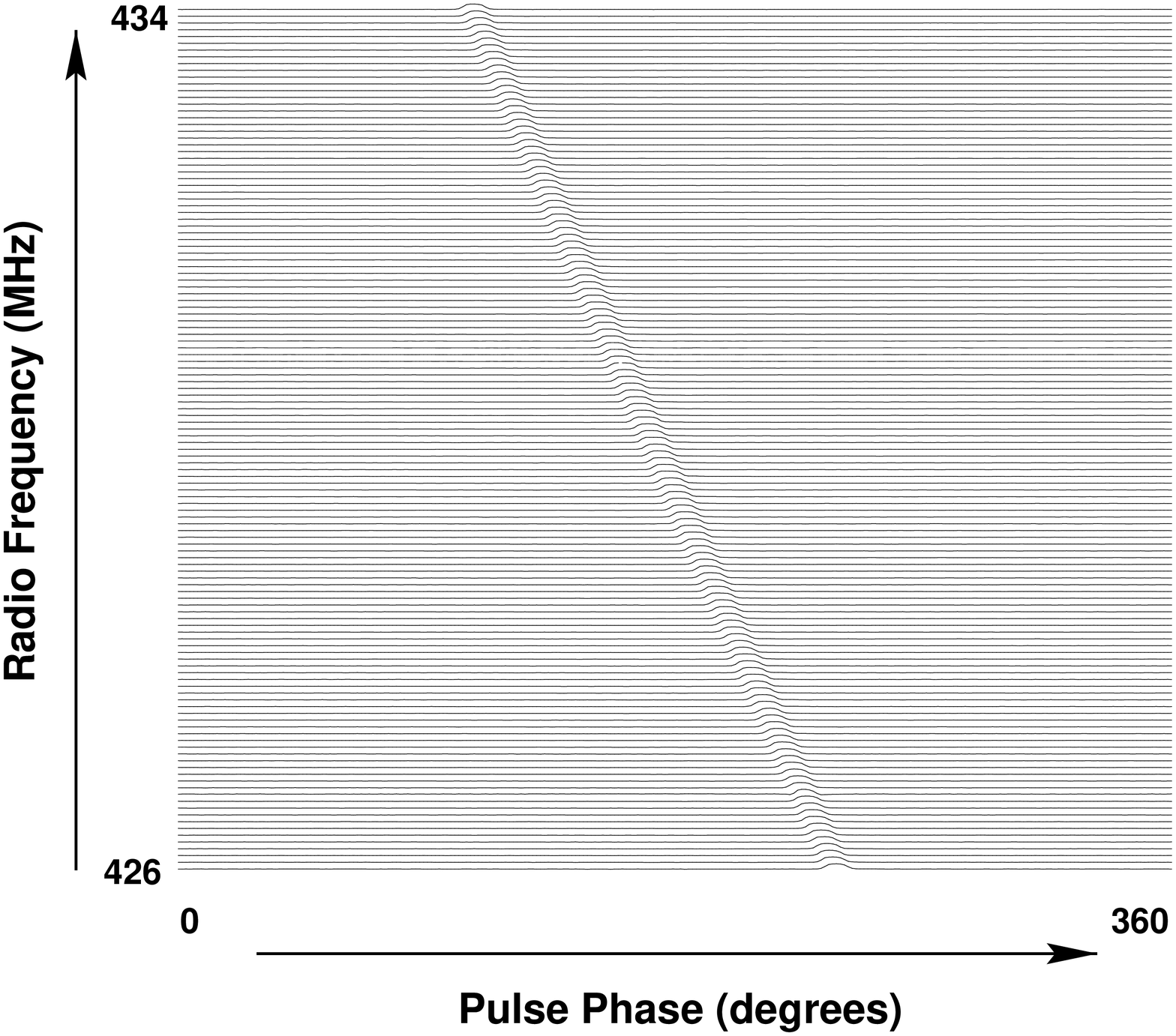}}
\end{picture}
\caption[]
{
Pulse dispersion shown in this recent 30-s Arecibo observation of PSR
B1933+16 across an 8-MHz passband centred at 430 MHz. The period of
this pulsar is 358.7 ms. The dispersive time delay between 434 MHz and
426 MHz is 133 ms.
}
\label{fig:1933}
\end{figure}

One of the phenomena clearly noted in the discovery paper was pulse
dispersion --- pulses at higher radio frequencies arrive earlier at
the telescope than their lower frequency counterparts. An example of
this is shown in Fig.~\ref{fig:1933}.  Hewish et al.~correctly
interpreted the effect as the frequency dependence of group velocity
of radio waves as they propagate through the interstellar medium --- a
cold ionised plasma.

Applying standard plasma physics formulae, it can be shown (see
e.g.~Lyne \& Smith 1998) that the difference in arrival times
$\Delta t$ between a high frequency $\nu_{\rm hi}$ (MHz) and a low frequency
$\nu_{\rm lo}$ (MHz) is given by
\begin{equation}
\label{equ:defdt}
 \Delta t = 4.15 \times 10^6 \, \, {\rm ms} \, \,
 \times (\nu_{\rm lo}^{-2} - \nu_{\rm hi}^{-2})  
 \times {\rm DM},
\end{equation}
where the dispersion measure DM (cm$^{-3}$ pc) is the integrated
column density of free electrons along the line of sight:
\begin{equation}
\label{equ:defdm}
{\rm DM} = \int_{\rm 0}^{d} \,\, n_{\rm e} \,\, dl.
\end{equation}
Here, $d$ is the distance to the pulsar (pc) and $n_{\rm e}$ is the
free electron density (cm$^{-3}$).  Pulsars at large distances have
higher column densities, and therefore larger DMs, than pulsars closer
to Earth so that, from Eq.~\ref{equ:defdt}, the dispersive delay
across the bandwidth is greater. In the original discovery paper,
Hewish et al.~(1968) measured a delay for the first pulsar (B1919+21)
of $\simeq 0.2$ s between 81.5 and 80.5 MHz.  From
Eq.~\ref{equ:defdt}, we infer a DM of about 13 cm$^{-3}$ pc.
Assuming, as a first-order approximation, that the mean Galactic
electron density is 0.03 cm$^{-3}$ (Ables \& Manchester 1976), this
implies a distance of about 0.4 kpc\footnote{Hewish et al.~(1968)
assumed 0.2 cm$^{-3}$ which resulted in an underestimated distance,
but still clearly demonstrated that the source is located well beyond
the solar system.}.

The most straightforward method to compensate for pulse dispersion is
to use a filterbank to sample the passband as a number of contiguous
\begin{figure}[hbt]
\setlength{\unitlength}{1in}
\begin{picture}(0,1.3)
\put(-.1,1.7){\includegraphics{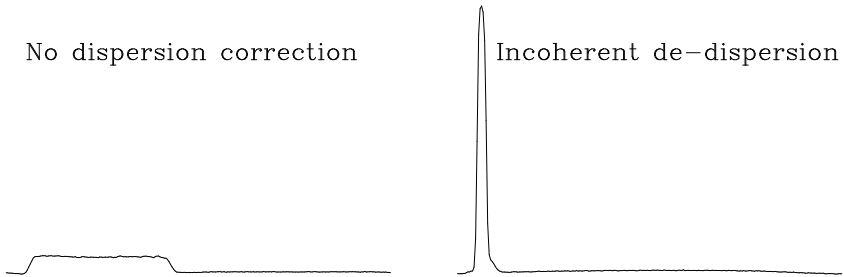}}
\end{picture}
\caption[]
{
Left: the ``raw profile'' of B1933+16 formed by the direct addition of
individual filterbank channels in Fig.~\ref{fig:1933}. Right:
incoherently de-dispersed profile formed by delaying the filterbank
channels before addition. Both displays have the same scale.
}
\label{fig:1933dis}
\end{figure}
channels and apply successively larger time delays (calculated from
Eq.~\ref{equ:defdt}) to higher frequency channels {\it before} summing
over all channels to produce a sharp ``de-dispersed'' profile.  This
can be carried out either in hardware or in software. Fig.~\ref{fig:1933dis} 
shows the clear gain in signal-to-noise ratio and time resolution achieved 
when the data from Fig.~\ref{fig:1933} are properly de-dispersed, as 
opposed to a simple detection over the whole bandwidth.

\begin{figure}[hbt]
\setlength{\unitlength}{1in}
\begin{picture}(0,2.7)
\put(0.4,-0.1){\includegraphics{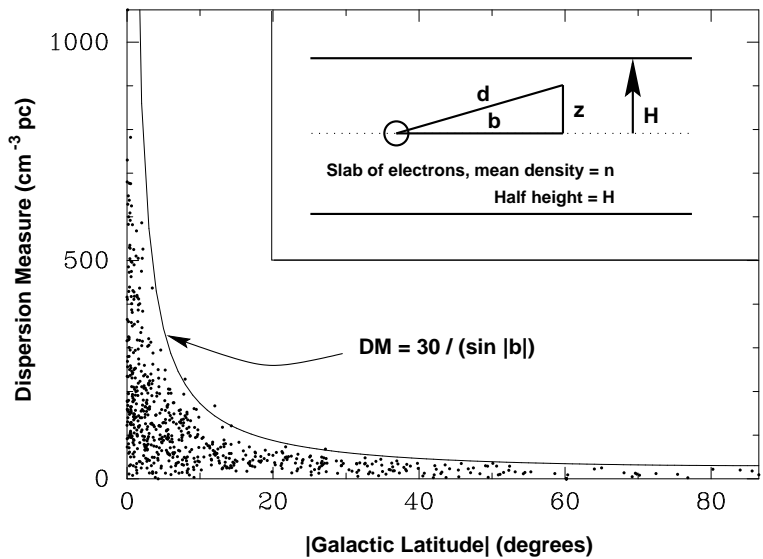}}
\end{picture}
\caption[]
{
Dispersion measures plotted against Galactic latitudes. Inset:
a simple slab model to explain the envelope of points in terms
of a finite electron layer (solid curve).
}
\label{fig:slab}
\end{figure}

The fact that the free electrons in the Galaxy are finite in extent is
well demonstrated by Fig.~\ref{fig:slab} which shows the dispersion
measures of 700 pulsars plotted against the absolute value of their
respective Galactic latitudes. It is straightforward to show that, for
a simple slab model of free electrons with a mean density $n$ and
half-height $H$, the maximum DM for a given line of sight along a
latitude $b$ is $H n/ \sin b$. The solid curve in Fig.~\ref{fig:slab}
shows fairly convincingly that this simple model accounts for the
trend rather well implying that $Hn\sim 30$ pc cm$^{-3}$.  Taking
$n=0.03$ cm$^{-3}$ as before gives us a first-order estimate of the
thickness of the electron layer --- about 1 kpc.

Independent measurements of pulsar distances can, for a large enough
sample, be fed back into Eq.~\ref{equ:defdm} to calibrate the Galactic
distribution of free electrons. There are three basic distance
measurement techniques: neutral hydrogen absorption, trigonometric
parallax (measured either with an interferometer or through pulse
time-of-arrival techniques) and from associations with objects of
known distance (i.e.~supernova remnants, globular clusters and
the Magellanic Clouds). Together, these provide measurements of (or
limits on) the distances to over 100 pulsars.  Taylor \& Cordes (1993)
have used these distances, together with measurements of interstellar
scattering for various Galactic and extragalactic sources, to
calibrate an electron density model. In a statistical sense, the model
can be used to provide distance estimates with an uncertainty of
$\sim$ 30\%. Although the model is free of large systematic trends,
its use to estimate distances to individual pulsars may result in
systematic errors by as much as a factor of two.

\section{Erratic Individual Pulses and Stable Integrated Profiles}
\label{sec:profs}

\begin{figure}[hbt]
\setlength{\unitlength}{1in}
\begin{picture}(0,3.6)
\put(0.05,-0.1){\includegraphics{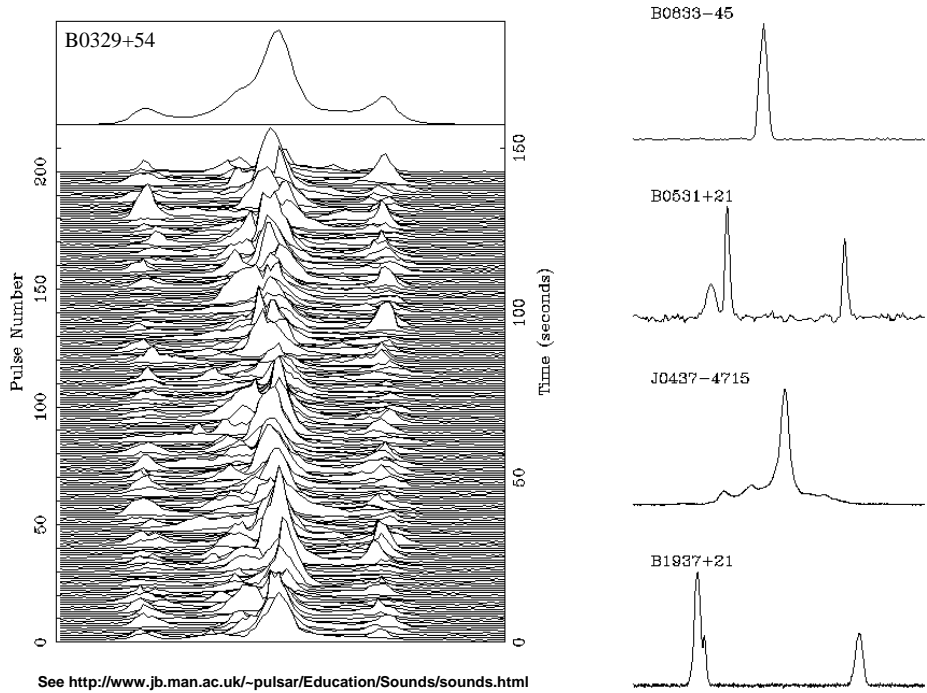}}
\end{picture}
\caption[]
{
Left: a sequence of 200 individual pulses received from the strong
pulsar B0329+54 showing the rich diversity of emission from one pulse
to the next. The sum of all the pulses forms a characteristic
``integrated profile'' shown for this pulsar in the box at the
top. Right: some other examples of these stable waveforms.
}
\label{fig:single}
\end{figure}

Pulsars are weak radio sources. Mean flux densities, usually quoted in
the literature at a radio frequency of 400 MHz, vary between 1 and 100
mJy (1 Jy $= 10^{-26}$ W m$^{-2}$ Hz$^{-1}$).  This means that the
addition of many thousands of pulses is required in order to produce a
discernible profile. Only a handful of sources presently known are
strong enough to allow studies of individual pulses.  A remarkable
fact from these studies is that, although the individual pulses vary
quite dramatically, at any particular observing frequency the
integrated profile is very stable. This is illustrated in 
Fig.~\ref{fig:single}.  

In the above examples of stable pulse profiles, which have been
normalised to represent 360 degrees of rotational phase, the astute
reader will notice two examples of so-called interpulses --- a
secondary pulse separated by about 180 degrees from the main
pulse. The most natural interpretation for this phenomenon is that the
two pulses originate from opposite magnetic poles of the neutron star
(see however Manchester \& Lyne 1977). Geometrically speaking, this is
a rather unlikely situation. As a result, the fraction of the known
pulsars with interpulses is only a few percent.

\begin{figure}[hbt]
\setlength{\unitlength}{1in}
\begin{picture}(0,1.5)
\put(0.15,0){\includegraphics{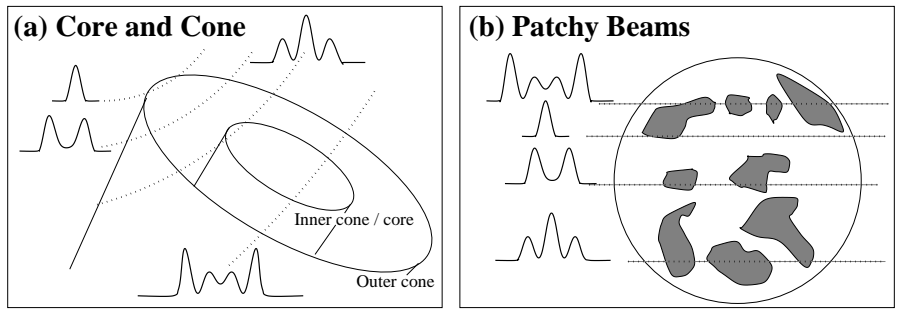}}
\end{picture}
\caption[]
{
Phenomenological models for pulse shape morphology produced
by different line-of-sight cuts of the beam. (Figure designed
by M.~Kramer and A.~von Hoensbroech).
}
\label{fig:shapes}
\end{figure}

The integrated pulse profile should really be thought of as a unique
``fingerprint'' of the radio emission beam of each neutron star.  The rich
variety of pulse shapes can be attributed to different line-of-sight
cuts through the radio beam of the neutron star as it sweeps past the
Earth.  Two contrasting phenomenological models which account for this
are shown in Fig.~\ref{fig:shapes}. The ``core and cone'' model,
proposed by Rankin (1983), depicts the beam as a core surrounded by a
series of nested cones. Alternatively, the ``patchy beam'' model,
championed by Lyne \& Manchester (1988), has the beam populated by a
series of emission regions.

\section{Pulsar Timing Basics}
\label{sec:timing}

Soon after their discovery, it became clear that pulsars are excellent
celestial clocks. Hewish et al.~(1968) demonstrated that the period of
the first pulsar, B1919+21, was stable to one part in $10^7$ over a
time-scale of a few months. Following the discovery of the millisecond
pulsar, B1937+21, in 1982 (Backer et al.~1982) it was demonstrated that
its period could be measured to one part in $10^{13}$ or better (Davis
et al.~1985). This unrivaled stability leads to a host of applications
including time keeping, probes of relativistic gravity and natural
gravitational wave detectors. Subsequently, a whole science has
developed to accurately measure the pulse time-of-arrival 
in order to extract as much information about each pulsar as possible.

\begin{figure}[hbt]
\setlength{\unitlength}{1in}
\begin{picture}(0,1.5)
\put(0.2,-0.1){\includegraphics{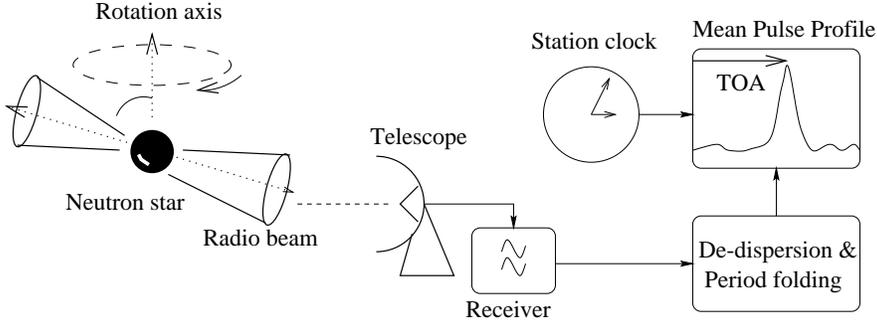}}
\end{picture}
\caption[]
{
Schematic showing the main stages involved in pulsar timing observations.
}
\label{fig:timing}
\end{figure}

Fig.~\ref{fig:timing} summarises the essential steps involved in a
pulse ``time-of-arrival'' (TOA) measurement. Incoming pulses emitted
by the rotating neutron star traverse the interstellar medium before
being received by the radio telescope. After amplification by high
sensitivity receivers, the pulses are de-dispersed (\S \ref{sec:dm})
and added to form a mean pulse profile.  During the observation, the
data regularly receive a time stamp, usually based on a maser at the
observatory, plus a signal from the GPS (Global Positioning System of
satellites) time system. The TOA of this mean pulse is then defined as
the arrival time of some fiducial point on the profile. Since the mean
profile has a stable form at any given observing frequency (\S
\ref{sec:profs}) the TOA can be accurately determined by a simple
cross-correlation of the observed profile with a high signal-to-noise
``template'' profile --- obtained from the addition of many
observations of the pulsar at a particular observing frequency.

In order to properly model the rotational behaviour of the neutron
star, we require TOAs as measured by an inertial observer.  Due to the
Earth's orbit around the Sun, an observatory located on Earth
experiences accelerations with respect to the neutron star. The
observatory is therefore not in an inertial frame. To a very good
approximation, the centre-of-mass of the solar system, the solar
system barycentre, can be regarded as an inertial frame. It is
standard practice to transform the observed TOAs to this frame using a
planetary ephemeris.

Following the accumulation of about ten to twenty barycentric TOAs
from observations spaced over at least several months, a surprisingly
simple model can be applied to the TOAs and optimised so that it is
sufficient to account for the arrival time of any pulse emitted during
the time span of the observations, and predict the arrival times of
subsequent pulses. The model is based on a Taylor expansion of the
angular rotational frequency about a model value at some reference
epoch to calculate a model pulse phase as a function of time. Based
upon this simple model, and using initial estimates of the position,
dispersion measure and pulse period, a ``timing residual'' is
calculated for each TOA as the difference between the observed and
predicted pulse phases (see e.g.~Lyne \& Smith 1998).

\begin{figure}[hbt]
\setlength{\unitlength}{1in}
\begin{picture}(0,2.7)
\put(-.1,-.85){\includegraphics{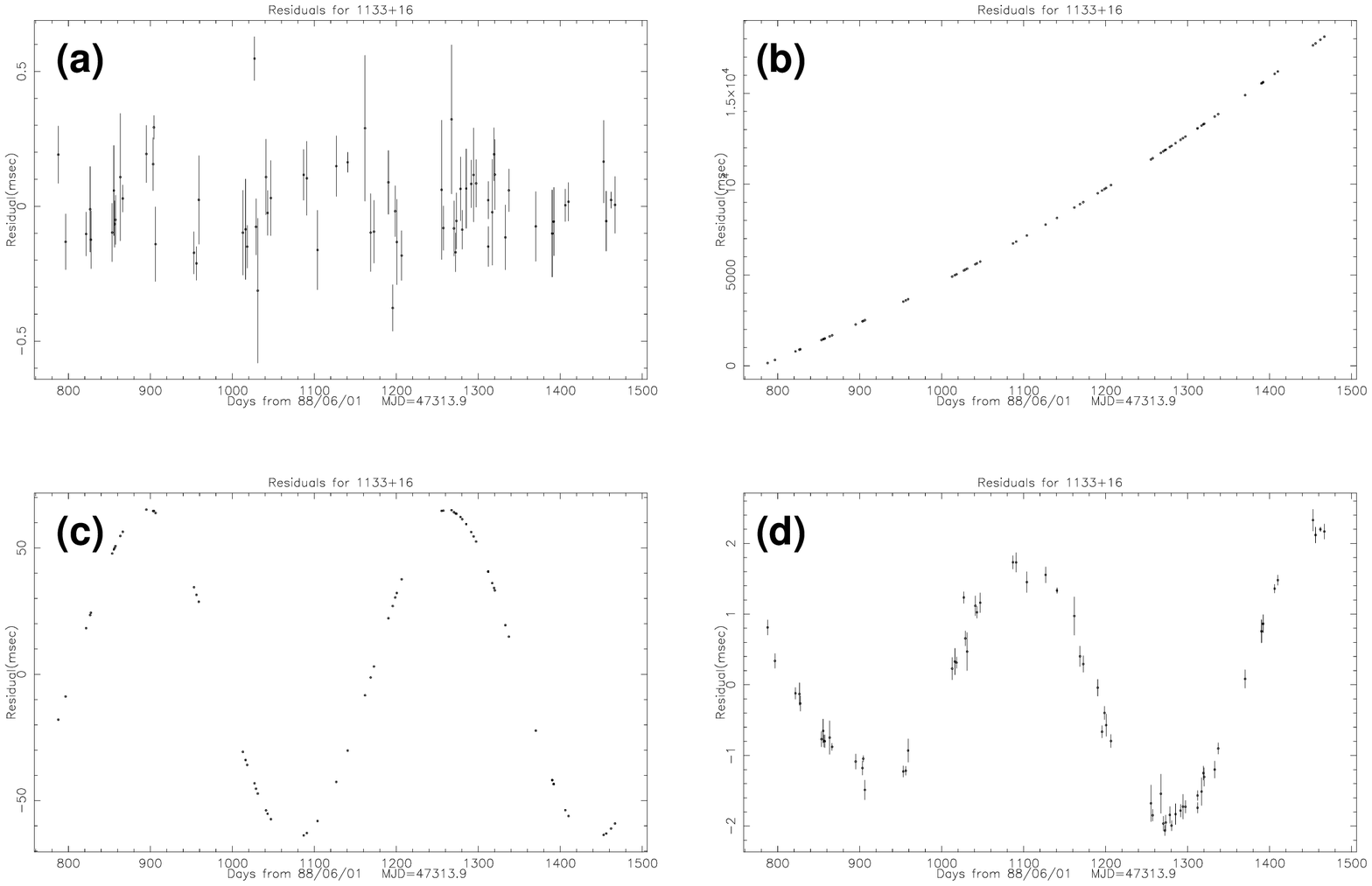}}
\end{picture}
\caption[]
{
Timing model residuals versus date for PSR B1133+16. Case (a) shows
the residuals obtained from the best fitting model which includes
period, period derivative, position and proper motion. Case (b) is the
result of setting the period derivative term to zero in this model.
Case (c) shows the effect of a 1 arcmin error in the assumed
declination.  Case (d) shows the residuals obtained assuming zero
proper motion.
}
\label{fig:1133}
\end{figure}

Ideally, the residuals should have a zero mean and be free from any
systematic trends (Fig.~\ref{fig:1133}a). Inevitably, however, due to
our {\it a-priori} ignorance of the rotational parameters, the model
needs to be refined in a bootstrap fashion.  Early sets of residuals
will exhibit a number of trends indicating a systematic error in one
or more of the model parameters, or a parameter not initially
incorporated into the model. For example, a parabolic trend results
from an error in the period time derivative (Fig.~\ref{fig:1133}b).
Additional effects will arise if the assumed position of the pulsar
used in the barycentric time calculation is incorrect. A position
error of just one arcsecond results in an annual sinusoid
(Fig.~\ref{fig:1133}c) with a peak-to-peak amplitude of about 5 ms for
a pulsar on the ecliptic; this is easily measurable for typical TOA
uncertainties of order one milliperiod or better.
Similarly, the effect of a proper motion produces an annual sinusoid
of linearly increasing magnitude (Fig.~\ref{fig:1133}d). Proper
motions are, for many long-period pulsars, more difficult to measure
due to the confusing effects of timing noise (a random walk process
seen in the timing residuals; see e.g.~Cordes \& Helfand 1980). For
these pulsars, interferometric techniques can be used to obtain proper
motions (see Ramachandran's contribution and references therein).

In summary, a phase-connected timing solution obtained over an
interval of a year or more will, for an isolated pulsar, provide an
accurate measurement of the period, the rate at which the neutron star
is slowing down, and the position of the neutron star.  Presently,
measurements of these essential parameters are available for about 600
pulsars. The implications of these measurements for the ages and
magnetic fields of neutron stars will be discussed in my other
contribution to this volume.

\begin{figure}[hbt]
\setlength{\unitlength}{1in}
\begin{picture}(0,1.4)
\put(-0.1,1.6){\includegraphics{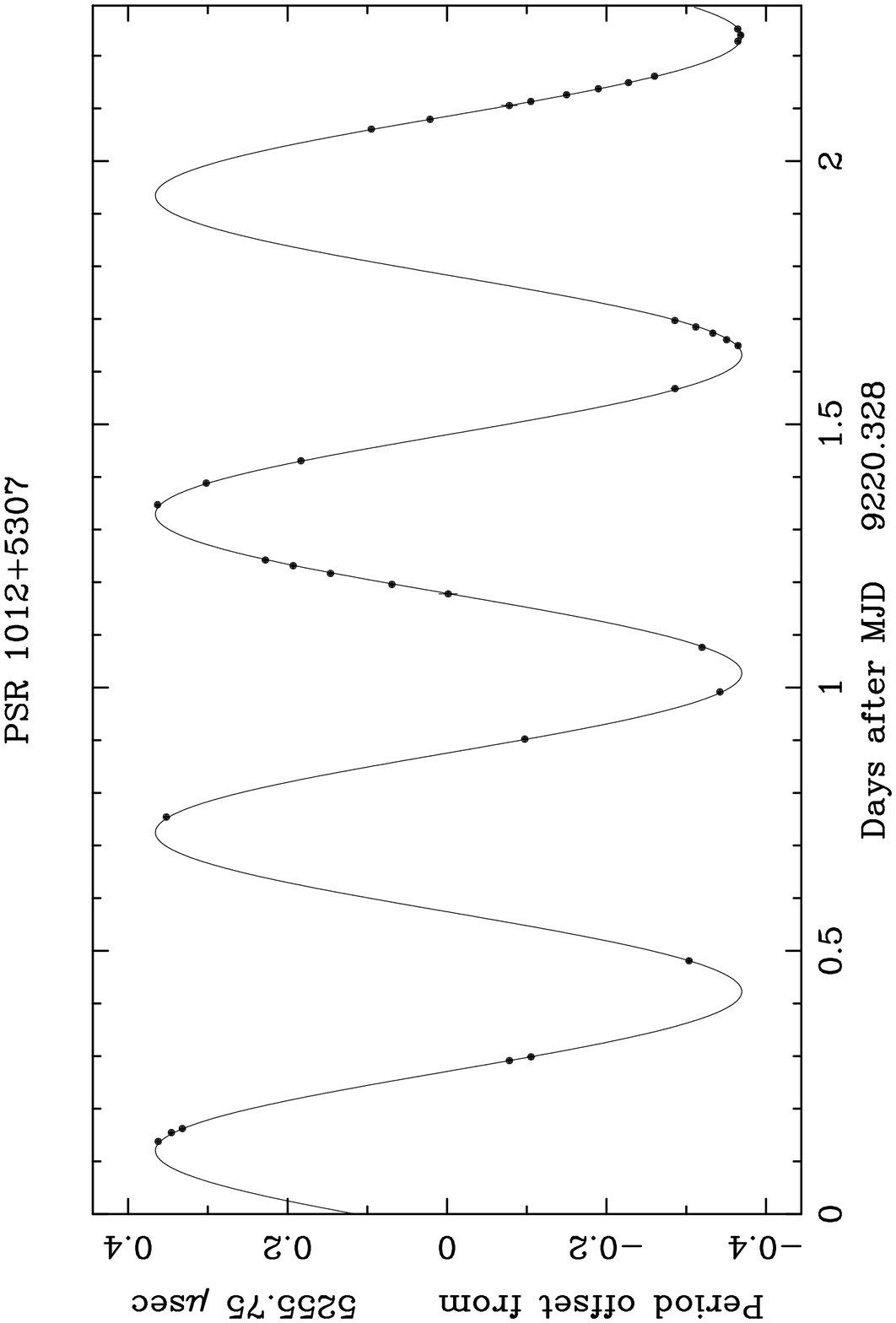}}
\put(+2.4,1.6){\includegraphics{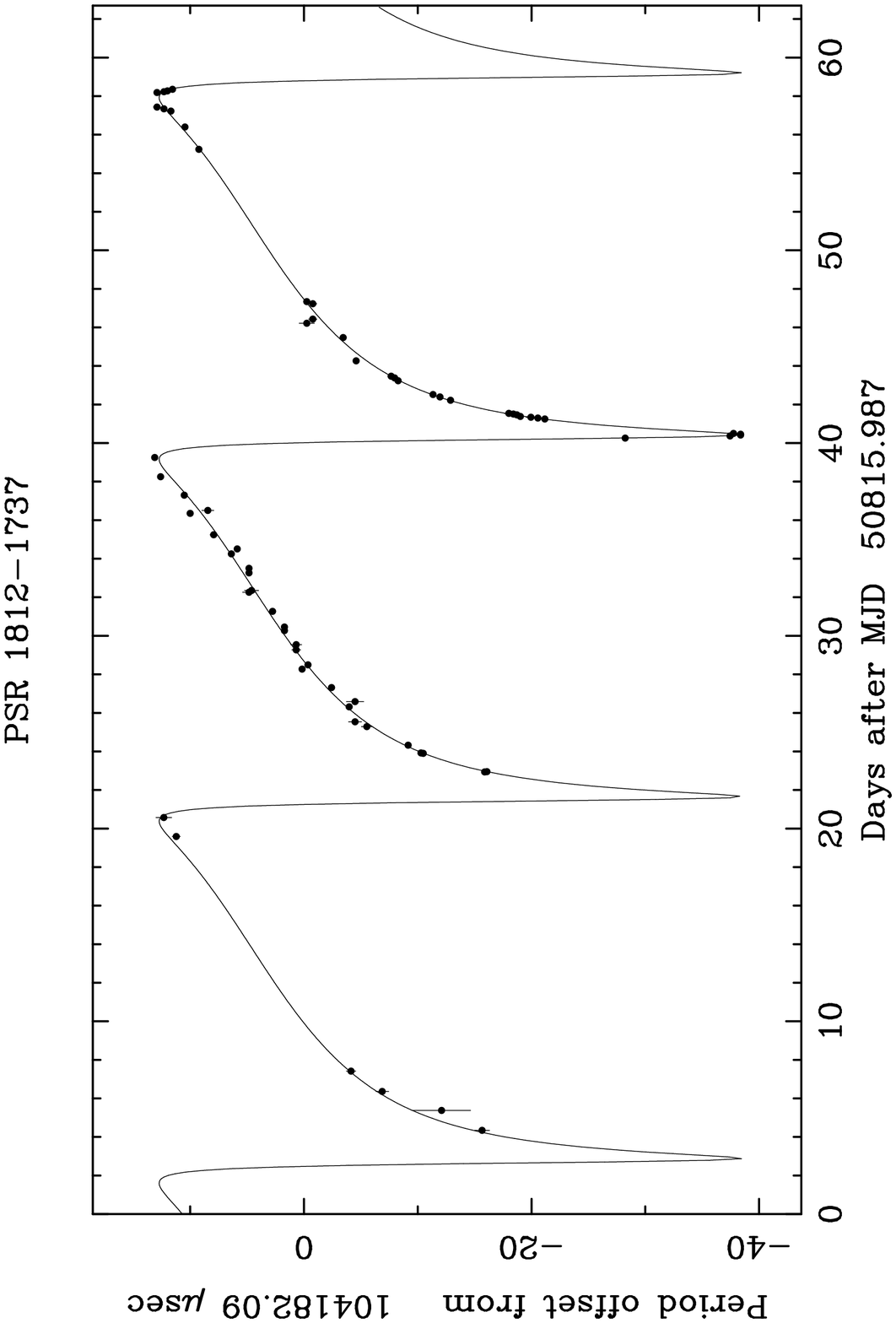}}
\end{picture}
\caption[] 
{
Two different binary pulsars. Left: PSR J1012$+$5307 --- a 5.25-ms
pulsar in a 14.5-hour circular orbit around a low-mass white dwarf
companion (Nicastro et al.~1995). Right: PSR J1811$-$1736 (Lyne et
al.~2000). A 104-ms pulsar in a highly eccentric 18.8-day orbit around
a massive companion (probably another neutron star).
}
\label{fig:orbits}
\end{figure}

For binary pulsars, the model needs to be extended to incorporate the
additional radial accelerations of the pulsar as it orbits the common
centre-of-mass of the binary system.  Treating the binary orbit using
just Kepler's laws to refer the TOAs to the binary barycentre requires
five additional model parameters: the orbital period, projected
semi-major orbital axis, orbital eccentricity, longitude of periastron
and the epoch of periastron passage.  The Keplerian description of the
orbit is identical to that used for spectroscopic binary stars where a
characteristic orbital ``velocity curve'' shows the radial component
of the star's velocity as a function of time. The analogous plot for
pulsars is the apparent pulse period against time.  For circular
orbits the behaviour is sinusoidal whilst for eccentric orbits the
curve has a ``saw-tooth'' appearance.  Two examples are shown in
Fig.~\ref{fig:orbits}.

Also by analogy with spectroscopic binaries, constraints on the mass
of the orbiting companion can be placed by combining the projected
semi-major axis $a_{\rm p}\,\sin\,i$ and the orbital period $P_{\rm
o}$ to obtain the mass function:
\begin{equation}
\label{equ:massfn}
f(m_{\rm p},m_{\rm c}) = \frac{4\pi^2}{G} \frac{(a_{\rm
p}\,\sin\,i)^3}{P_{\rm o}^2} = \frac{(m_{\rm c}\,\sin\,i)^3}{(m_{\rm
p}+m_{\rm c})^2},
\end{equation}
where $G$ is the universal gravitational constant.  Assuming a pulsar
mass $m_{\rm p}$ of 1.35 M$_{\odot}$ (see below), the mass of the
orbiting companion $m_{\rm c}$ can be estimated as a function of the
(initially unknown) angle $i$ between the orbital plane and the plane
of the sky. The minimum companion mass $m_{\rm min}$ occurs when the
orbit is assumed edge-on ($i=90^{\circ}$).

Further information on the orbital inclination and component masses
may be obtained by studying binary systems which exhibit a number of
relativistic effects not described by Kepler's laws. Up to five 
``post-Keplerian'' parameters exist within the framework of general
relativity. Three of these parameters (the rate of periastron advance,
a gravitational redshift parameter, and the orbital period derivative)
have been measured for the original binary pulsar, B1913+16 (Taylor \&
Weisberg 1989), allowing high-precision measurements of the masses of both
components, as well as stringent tests of general relativity.
A further two post-Keplerian parameters related to the Shapiro
delay in the double neutron star system PSR B1534+12 have now also
been measured (Stairs et al.~1998).  Based on these results, and other
radio pulsar binary systems, Thorsett \& Chakrabarty (1999) have
recently demonstrated that the range of neutron star masses has a
remarkably narrow underlying Gaussian distribution with a mean of 1.35
M$_{\odot}$ and a standard deviation of only 0.04 M$_{\odot}$.

\section{Pulsar Searching}
\label{sec:searching}

Pulsar searching is, conceptually at least, a rather simple process
--- the detection of a dispersed, periodic signal hidden in a noisy
time series data taken with a radio telescope. In what follows we give
a only brief description of the basic search techniques.  Further
discussions can be found in Lyne (1988), Nice (1992) and Lorimer
(1998) and references therein.

\begin{figure}[hbt]
\setlength{\unitlength}{1in}
\begin{picture}(0,2.4)
\put(0.1,-0.08){\includegraphics{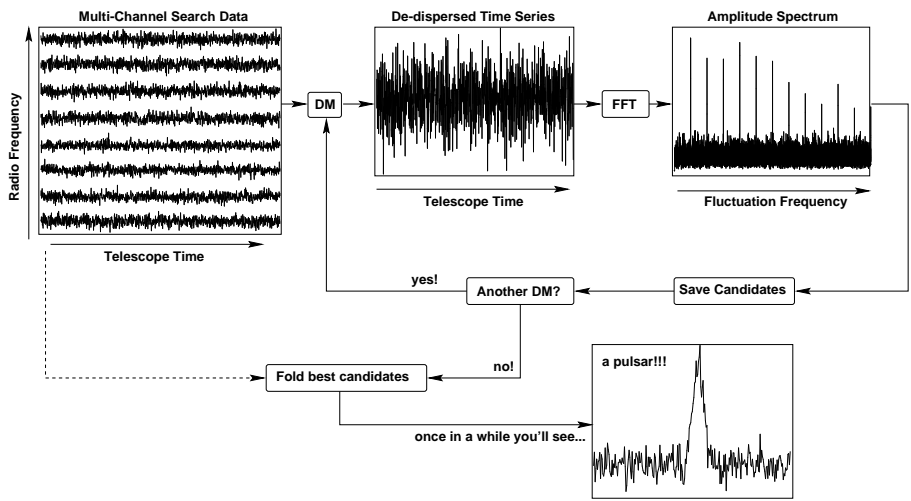}}
\end{picture}
\caption[]
{
Schematic summarising the essential steps in a ``standard'' pulsar search.
}
\label{fig:search}
\end{figure}

Most pulsar searches can be pictured as shown in
Fig.~\ref{fig:search}. The multi-channel search data is typically
collected using a filterbank or a correlator (see e.g.~Backer et
al.~1990), either of which usually provides a much finer
channelisation than the eight channels shown for illustrative purposes
in Fig.~\ref{fig:search}.  The channels are then incoherently
de-dispersed (see \S \ref{sec:dm}) to form a single noisy time
series. An efficient way to find a periodic signal in these data is to
take the Fast Fourier Transform (FFT) and plot the resulting amplitude
spectrum. For a narrow duty cycle the spectrum will show a family of
harmonics which show clearly above the noise. To detect weaker signals
still, a harmonic summing technique is usually implemented at this stage
(see e.g.~Lyne 1988). The best candidates are then saved and the whole 
process is repeated for another trial DM.

\begin{figure}[hbt]
\setlength{\unitlength}{1in}
\begin{picture}(0,2.78)
\put(0.1,3.4){\includegraphics{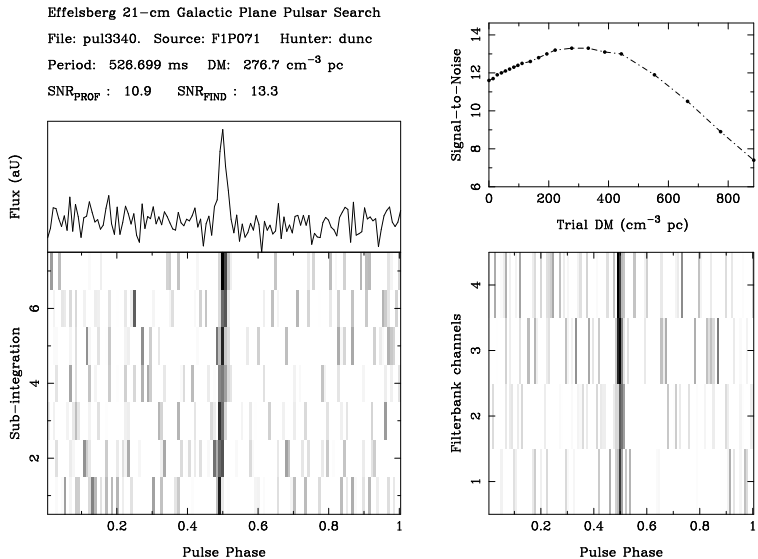}}
\end{picture}
\caption[]
{
Example search code output showing PSR J1842--0415 --- the first
pulsar ever discovered with the 100-m Effelsberg radio telescope
(Lorimer et al.~2000). 
}
\label{fig:epsr}
\end{figure}

After a sufficiently large number of DMs have been processed, a list
of pulsar candidates is compiled and it is then a matter of folding
the raw time series data at the candidate period.  Fig.~\ref{fig:epsr}
is an excellent example of the characteristics of a strong pulsar
candidate. The high signal-to-noise integrated profile (top left
panel) can be seen as a function of time and radio frequency in the
grey scales (lower left and right panels).  In addition, the dispersed
nature of the signal is immediately evident in the upper right hand
panel which shows the signal-to-noise ratio as a function of trial
DM. This combination of diagnostics proves extremely useful in
differentiating between pulsar candidates and spurious interference.

In the discussion hitherto we have implicitly assumed that the
apparent pulse period remains constant throughout the observation. For
searches with long integration times (Fig.~\ref{fig:epsr} represents a
35-min observation), this assumption is only valid for solitary
pulsars, or those in binary systems where the orbital periods are
longer than about a day. For shorter-period binary systems, as noted
by Johnston \& Kulkarni (1992), the Doppler shifting of the period
results in a spreading of the total signal power over a number of
frequency bins in the Fourier domain. Thus, a narrow harmonic becomes
smeared over several spectral bins.

To quantify this effect, consider the resolution in fluctuation
frequency (the width of a Fourier bin) $\Delta f = 1/T$, where $T$ is
the length of the integration. It is straightforward to show that the
drift in frequency of a signal due to a constant
acceleration\footnote{The smearing is even more severe if $a$ varies
--- i.e.~extremely short orbital periods.}  $a$ during this time is
$aT/(Pc)$, where $P$ is the true period of the pulsar and $c$ is the
speed of light.  Comparing these two quantities, we note that the
signal will drift into more than one spectral bin if $aT^2/(Pc) >
1$. Thus, without due care, long integration times potentially kill
off all sensitivity to short-period pulsars in exciting tight orbits
where the line-of-sight accelerations are high!

\begin{figure}[hbt]
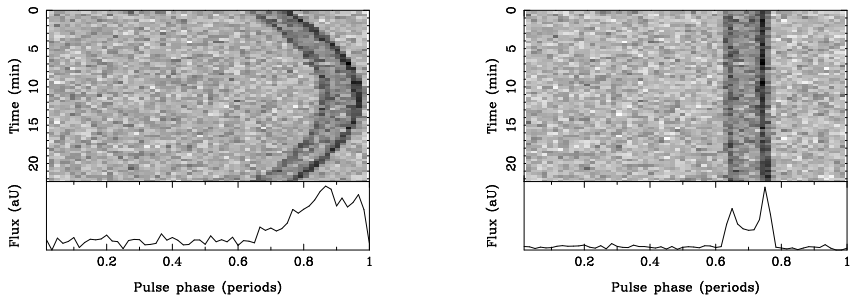

\setlength{\unitlength}{1in}
\begin{picture}(0,1.4)
\put(0.0,1.8){\includegraphics{1913_raw.ps}}
\put(2.5,1.8){\includegraphics{1913_cor.ps}}
\end{picture}
\caption[] 
{
Left: a 22.5-min Arecibo observation of the binary pulsar B1913+16.
The assumption that the pulsar has a constant period during this time
is clearly inappropriate given the drifting in phase of the pulse
during the integration (grey scale plot). Right: the same observation
after applying an acceleration search. This shows the effective
recovery of the pulse shape and a significant improvement (factor of
7) in the signal-to-noise ratio.
}
\label{fig:1913}
\end{figure}

As an example of this effect, as seen in the time domain,
Fig.~\ref{fig:1913} shows a 22.5-min search mode observation of
Hulse \& Taylor's binary pulsar B1913+16. Although this observation
covers only about 5\% of the orbit (7.75 hr), the effects of the
Doppler smearing on the pulse signal are very apparent. Whilst
the search code nominally detects the pulsar with a signal-to-noise
ratio of 9.5 for this observation, it is clear that the Doppler
shifting of the pulse period seen in the individual sub-integrations
results in a significant reduction in the signal-to-noise ratio.

Pulsar searches of distant globular clusters are most prone to this
effect since long integration times are required to reach a reasonable
level of sensitivity. Since one of the motivations for searching
clusters is their high specific incidence of low-mass X-ray binaries,
it is likely that short orbital period pulsars will also be
present. Anderson et al.~(1990) were the first to really address
this problem during their survey of a number of globular clusters
using the Arecibo radio telescope. In the so-called acceleration
search, the pulsar is assumed to have a constant acceleration ($a$)
during the integration.  Each de-dispersed time series can then be
re-sampled to refer it to the frame of an observer with an identical
acceleration.  This transformation is readily achieved by applying the
Doppler formula to relate a time interval in the pulsar frame, $\tau$,
to that in the observed frame at time $t$, as $\tau(t) = \tau_0 ( 1 +
at/c )$, where $a$ is the observed radial acceleration of the pulsar
along the line-of-sight, $c$ is the speed of light, and $\tau_0$ is a
normalising constant (for further details, see Camilo et al.~2000a). If
the correct acceleration is chosen, then the net effect is a time
series containing a signal with a constant period which can be found
using the standard pulsar search outlined above. An example of this is
shown in the right panel of Fig.~\ref{fig:1913} where the time series
has been re-sampled assuming a constant acceleration of --17 m s$^{-2}$.
The signal-to-noise ratio is increased to 67!

The true acceleration is, of course, {\it a-priori} unknown, meaning
that a large number of acceleration values must be tried in order to
``peak up'' on the correct value.  Although this necessarily adds an
extra dimension to the parameter space searched it can pay handsome
dividends, particularly in globular clusters where the dispersion
measure is well constrained by pulsars already discovered in the same
cluster. Anderson et al.~(1990) used this technique to find PSR
B2127+11C --- a double neutron star binary in M15 which has parameters
similar to B1913+16. Camilo et al.~(2000a) have recently applied the
same technique to 47~Tucanae, a globular cluster previously known to
contain 11 millisecond pulsars, to aid the discovery of a further 9
binary millisecond pulsars in the cluster.

The new discoveries in 47~Tucanae include a 3.48-ms pulsar in 96-min
orbit around a low-mass companion (Camilo et al.~2000a). Whilst this is
presently the shortest orbital period for a radio pulsar binary, the 
mere existence of this pulsar, as well as the 11-min X-ray binary
X1820$-$303 in the globular cluster NGC6624 (Stella et al.~1987),
strongly suggests that there is a population of extremely short-period
radio pulsar binaries residing in globular clusters just waiting to be
found.  As Camilo et al.~demonstrate, the assumption of a constant
acceleration during the observation clearly breaks down for such short
orbital periods, requiring alternative techniques.

One obvious extension is to include a search over the time derivative
of the acceleration. This is currently being tried on some of the
47~Tucanae data. Although this does improve the sensitivity to
short-period binaries, it is computationally rather costly. An
alternative technique developed by Ransom, Cordes \& Eikenberry (in
prep.~see also astro-ph/9911073) looks to be particularly efficient at
finding binaries whose orbits are so short that many orbits can take
place during an integration. This {\it phase modulation
technique} exploits the fact that the periodic signals from such a
binary are modulated by the orbit to create a family of periodic
sidebands around the nominal spin period of the pulsar. This technique
appears to be extremely promising and is currently being applied to
radio and X-ray search data.

\section{Recent Survey Highlights --- The Parkes Multibeam Survey}
\label{sec:recent}

No current review on pulsar searching would be complete without
summarising the revolution in the field that is presently taking place
at the 64-m Parkes radio telescope in New South Wales, Australia.
With $13 \times 20$-cm 25-K receivers on the sky, along with $13
\times 2 \times 288$-MHz filterbanks, the telescope is presently
making major contributions in a number of different pulsar search
projects. In its main use for a Galactic plane survey (Camilo et
al.~2000b), the system achieves a sensitivity of 0.15 mJy in 35 min
and covers about one square degree of sky per hour of observing --- a
standard that is far beyond the present capabilities of any other
observatory.

The staggering total of over 500 new pulsars has come from an analysis
of about half the total data. Such a large haul is resulting in
significant numbers of interesting individual objects: several of the
new pulsars are observed to be spinning down at high rates, suggesting
that they are young objects with large magnetic fields. The inferred
age for the 400-ms pulsar J1119$-$6127, for example, is only 1.6
kyr. Another member of this group is the 4-s pulsar J1814$-$1744, an
object that may fuel the ever-present ``injection'' controversy
surrounding the initial spin periods of neutron stars (see however my
other contribution to these proceedings).

A number of the new discoveries from the survey have orbiting
companions. Several low-eccentricity systems are known where the
likely companions are white dwarf stars. Two possible double neutron
star systems are presently known: J1811$-$1736 (see
Fig.~\ref{fig:orbits}), while J1141$-$65 has a lower eccentricity but
an orbital period of only 4.75 hr. The fact that J1141$-$65 may have a
characteristic age of just over 1 Myr implies that the likely
birth-rate of such objects may be large. Although tempting, it is
premature to extrapolate the properties of one object. It is,
however, clear that these binary systems, and the many which will
undoubtably come from this survey, will teach us much about the still
poorly-understood population of double neutron star systems (see
Kalogera's contribution in this volume).

The Parkes multibeam system has not only been finding young, distant
pulsars along the Galactic plane. Edwards et al.~(in prep.~see also
astro-ph/9911221) have been using the same system to search
intermediate Galactic latitudes ($5^{\circ} \leq |b| \leq
15^{\circ}$). The discovery of 8 short-period pulsars during this
search, not to mention 50 long-period objects, strongly supports a
recent suggestion by Toscano et al.~(1998) that L-band ($\lambda$ 20
cm) searches are an excellent means of finding relatively distant
millisecond pulsars.

The most massive binary system yet from either of these two multibeam
surveys is J1740$-$3052, whose orbiting companion must be at least 11
M$_{\odot}$! Recent optical observations (Manchester et al.~2000)
reveal a K-supergiant as being the likely companion star in this
system. With such high-mass systems in the Galaxy, surely it is only a
matter of time before a radio pulsar will be found orbiting a
stellar-mass black hole.

Having finally made the connection between neutron stars and black
holes, I will finish by reiterating that pulsar astronomy is
currently enjoying one of the most productive phases in its
history. The new discoveries are sparking off a variety of follow-up
studies of all the exciting new objects. There will surely be plenty
of surprises in the coming years and new students are encouraged to
join this hive of activity.

\acknowledgements
Many thanks to Chris Salter and Fernando Camilo for their comments on
an earlier version of this manuscript. The Arecibo Observatory is
operated by Cornell University under a cooperative agreement with the NSF.

\end{document}